\documentclass[sigconf]{acmart}
\copyrightyear{2025}
\acmYear{2025}
\setcopyright{cc}
\setcctype{by}
\acmConference[UbiComp Companion '25]{Companion of the 2025 ACM International Joint Conference on Pervasive and Ubiquitous Computing}{October 12--16, 2025}{Espoo, Finland}
\acmBooktitle{Companion of the 2025 ACM International Joint Conference on Pervasive and Ubiquitous Computing (UbiComp Companion '25), October 12--16, 2025, Espoo, Finland}\acmDOI{10.1145/3714394.3756157}
\acmISBN{979-8-4007-1477-1/2025/10}

\usepackage{soul, xcolor}
\usepackage{subcaption}
\usepackage{graphicx}

\begin{document}

\title{Listening to the Mind: Earable Acoustic Sensing of Cognitive Load}

\author{Xijia Wei}
\email{xijia.wei.21@ucl.ac.uk}
\orcid{0000-0003-4745-6569}
\affiliation{%
  \department{UCL Interaction Centre}
  \institution{University College London}
  \city{London}
  \country{UK}
}
\authornote{These authors contributed equally.}

\author{Ting Dang}
\email{ting.dang@unimelb.edu.au}
\orcid{0000-0003-3806-1493}
\affiliation{%
  \department{School of Computing and Information Systems}
  \institution{University of Melbourne}
  \city{Melbourne}
  \country{Australia}
}
\authornotemark[1]
\authornote{Corresponding author.}

\author{Khaldoon Al-Naimi}
\email{khaldoon.al-naimi@nokia-bell-labs.com}
\orcid{0009-0009-5515-9987}
\affiliation{%
  \institution{Nokia Bell Labs}
  \city{Cambridge}
  \country{UK}
}

\author{Yang Liu}
\email{yang.16.liu@nokia-bell-labs.com}
\orcid{0000-0002-8452-6020}
\affiliation{%
  \institution{Nokia Bell Labs}
  \city{Cambridge}
  \country{UK}
}

\author{Fahim Kawsar}
\email{fahim.kawsar@nokia-bell-labs.com}
\orcid{0000-0001-5057-9557}
\affiliation{%
  \institution{Nokia Bell Labs}
  \city{Cambridge}
  \country{UK}
}

\author{Alessandro Montanari}
\orcid{0000-0003-4444-6242}
\email{alessandro.montanari@nokia-bell-labs.com}
\affiliation{%
  \institution{Nokia Bell Labs}
  \city{Cambridge}
  \country{UK}
}

\renewcommand{\shortauthors}{Xijia Wei et al.}

\begin{abstract}
% Monitoring cognitive load through auditory processing can help infer cognitive fatigue, enhance learning efficiency, and enable early detection of hearing impairments. With the rise of earable devices, this study is the first to explore the association between auditory cognitive load and perception in daily scenarios, using earable acoustic sensing. We: i) designed speech and audio tasks to induce varying cognitive loads; ii) developed stimulus to measure auditory characteristics; iii) proposed methods to quantify the correlation between auditory perception and cognitive load. Findings show a positive correlation between increased cognitive load and changes in auditory behaviour (p $<$ 0.01), with 63.2\% of participants showing higher sensitivity at 3 kHz. Gender and age-related differences were also observed. This study highlights the potential of earable acoustic sensing as a cost-effective method for cognitive load assessment, opening doors to broad future applications.

Earable acoustic sensing offers a powerful and non-invasive modality for capturing fine-grained auditory and physiological signals directly from the ear canal, enabling continuous and context-aware monitoring of cognitive states. As earable devices become increasingly embedded in daily life, they provide a unique opportunity to sense mental effort and perceptual load in real time through auditory interactions. In this study, we present the first investigation of cognitive load inference through auditory perception using acoustic signals captured by off-the-shelf in-ear devices. We designed speech-based listening tasks to induce varying levels of cognitive load, while concurrently embedding acoustic stimuli to evoke Stimulus Frequency Otoacoustic Emission (SFOAEs) as a proxy for cochlear responsiveness. Statistical analysis revealed a significant association (p < 0.01) between increased cognitive load and changes in auditory sensitivity, with 63.2\% of participants showing peak sensitivity at 3 kHz. Notably, sensitivity patterns also varied across demographic subgroups, suggesting opportunities for personalized sensing. Our findings demonstrate that earable acoustic sensing can support scalable, real-time cognitive load monitoring in natural settings, laying a foundation for future applications in augmented cognition, where everyday auditory technologies adapt to and support the user’s mental state.
\end{abstract}

\begin{CCSXML}
<ccs2012>
   <concept>
       <concept_id>10003120.10003138.10003140</concept_id>
       <concept_desc>Human-centered computing~Ubiquitous and mobile computing systems and tools</concept_desc>
       <concept_significance>500</concept_significance>
       </concept>
   <concept>
       <concept_id>10003120.10003121.10003128.10010869</concept_id>
       <concept_desc>Human-centered computing~Auditory feedback</concept_desc>
       <concept_significance>300</concept_significance>
       </concept>
   <concept>
       <concept_id>10003120.10003138.10003139.10010904</concept_id>
       <concept_desc>Human-centered computing~Ubiquitous computing</concept_desc>
       <concept_significance>500</concept_significance>
       </concept>
 </ccs2012>
\end{CCSXML}

\ccsdesc[500]{Human-centered computing~Ubiquitous computing}
\ccsdesc[500]{Human-centered computing~Ubiquitous and mobile computing systems and tools}
\ccsdesc[300]{Human-centered computing~Auditory feedback}

\keywords{auditory systems, cognitive load, earable sensing, acoustic analysis}


\maketitle

\section{Introduction}\label{introduction}
Imagine being in a bustling pub, trying to follow your friend’s voice across the room. This exemplifies the well-known “cocktail party” problem in auditory perception, where the challenge lies in focusing on a single speaker amid numerous competing sounds. While the ears receive all ambient noise, it is the brain’s cognitive processes that selectively filter and attend to the relevant auditory input. This selective attention requires substantial cognitive effort, as the brain must actively suppress distractions and allocate mental resources to manage the complexity of the auditory scene. Thus, listening in such environments involves an intricate interplay between sensory perception and cognitive load. %Understanding this relationship is critical because it highlights that auditory processing cannot be fully understood without considering the cognitive demands involved in real-world listening.
\begin{figure}[t]
\centering
\includegraphics[width=0.8\linewidth]{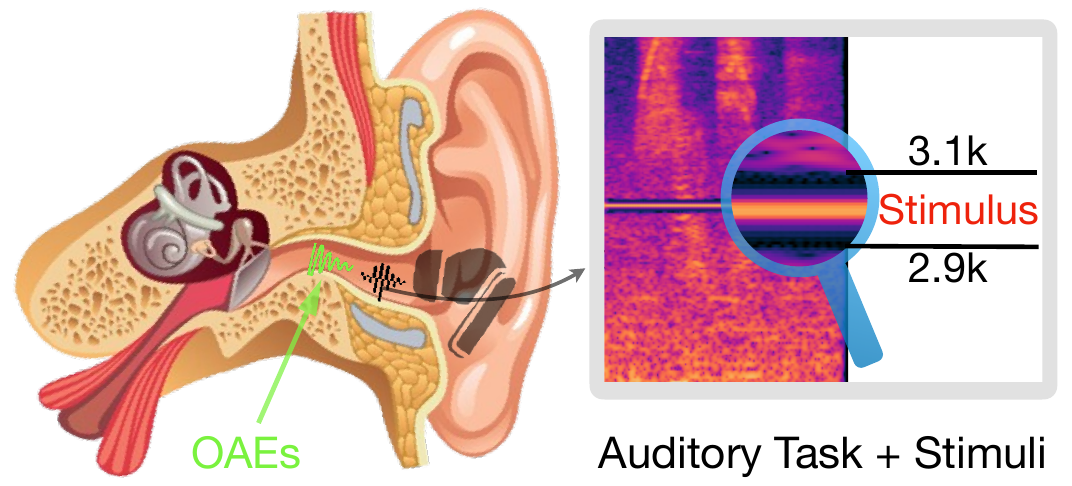}
% \vspace{-12pt}
\caption{The auditory task induces varying levels of cognitive load while simultaneously delivering acoustic stimuli to evoke otoacoustic emissions (OAEs). These emissions, measured via the earable device, reflect ear function and can be used to infer cognitive load. As shown in the example audio spectrum, both the task and stimulus are played concurrently. To enable accurate OAE detection, overlapping narrowband frequencies from the auditory task are removed, preserving only the stimulus components in the target band, leaving space for receiving OAE responses.}
%The auditory task is designed to induce varying levels of cognitive load, while simultaneously delivering acoustic stimuli that evoke otoacoustic emissions (OAEs). These OAEs provide measurable auditory responses that can be used to infer cognitive load based on changes in ear function. Given an example of an audio spectrum sent into the ear canal via the earable device, it illustrates that both the auditory task and the stimulus can be played simultaneously. While the auditory task elicits cognitive load, the stimulus is designed to evoke OAEs for measurement. To prevent signal overlap and ensure accurate OAE detection, narrowband frequencies in the auditory task that overlap with the stimulus are removed, leaving only the stimulus frequencies in the targeted band for OAE measurement.}
\vspace{-8pt}
\label{fig:concept}
\end{figure}
Understanding how cognitive load emerges during complex listening tasks is essential for designing technologies that respond to the mental and sensory demands of real-world interactions. Insights into this relationship have implications for learning efficiency, early detection of cognitive decline, and the development of personalised auditory interfaces that adapt to user state.

The increasing adoption of earable devices opens exciting new possibilities for in-situ sensing across a wide range of applications. While prior work has integrated EEG and photoplethysmography (PPG) sensors into earables~\cite{montanari2024omnibuds, ferlini2021ear, balaji2023stereo, romero2024optibreathe, mikkelsen2015eeg}, recent research has begun to explore the unique potential of in-ear microphones for acoustic sensing~\cite{butkow2024evaluation, liu2024respear, truong2022non, ma2024clearspeech, hu2023lightweight, demirel2024unobtrusive}.  Existing studies have demonstrated that in-ear microphones can capture subtle body-conducted sounds transmitted through the ear canal, facilitating health monitoring applications such as respiratory and cardiac signal detection~\cite{butkow2024evaluation, liu2024respear}. Notably, consumer-grade earbuds have even been repurposed for detecting hearing loss through acoustic sensing~\cite{chan2023wireless, shahid2024towards, demirel2023cancelling}. 

More importantly, this modality provides a direct window into auditory system dynamics and environmental interactions, enabling novel insights into hearing, cognitive load, and real-world listening contexts, an intersection of auditory sensing and cognitive state monitoring that remains largely unexplored with acoustic earable sensing. This is due in part to three key challenges:
i) designing auditory tasks that reliably induce measurable cognitive load,
ii) capturing auditory characteristics concurrently with playback under different cognitive load, and
iii) modelling the dynamic relationship between cognitive load and auditory response.

\begin{figure*}[t]
  \includegraphics[width=0.9\textwidth]{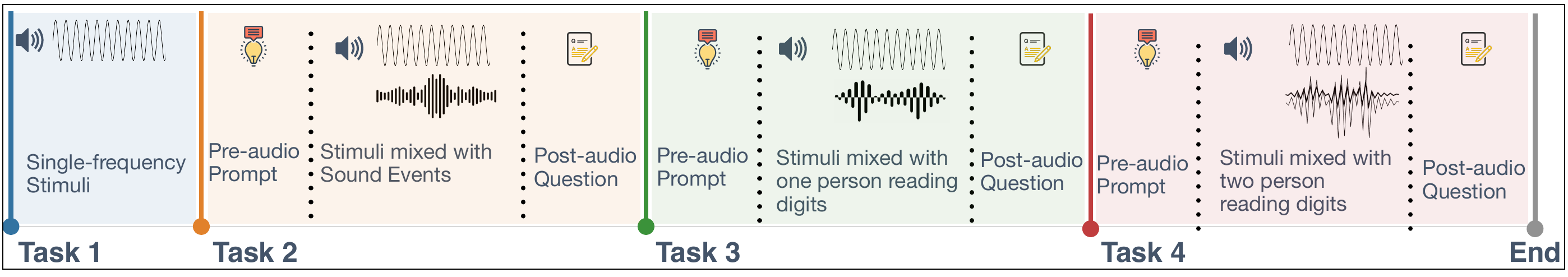}
  \caption{Auditory cognitive load task design, which contains four different activities to elicit progressively increasing levels of cognitive load.}
  \Description{Auditory cognitive load task design, which contains four different activities to elicit progressively increasing levels of cognitive load.}
  \vspace{-8pt}
  \label{fig:figure_illustation}
\end{figure*}

To address these challenges, our work proposes the first integrated framework that leverages earable acoustic sensing to capture auditory characteristics for cognitive load inference using off-the-shelf earable devices (Figure~\ref{fig:concept}). Our contributions include:
i) a set of ecologically valid auditory tasks that evoke varying cognitive load levels, validated through EEG recordings;
ii) the design of acoustic stimuli to evoke otoacoustic emissions (OAEs), enabling non-intrusive and real-time probing of auditory system response; and
iii) novel signal processing methods combined with statistical analyses to quantify the relationship between in-ear acoustic responses and cognitive load.

Our findings show that cognitive load induces measurable changes in earable-based acoustic sensing, and we further observe variation across demographic subgroups. This opens up new possibilities for everyday cognitive monitoring through acoustic sensing, leveraging the ear as both a channel for interaction and a sensor for mental state. By integrating cognitive state awareness into earable auditory systems, we aim to support richer, more adaptive, and context-aware auditory experiences, pushing the boundaries of how we use sound to sense, support, and enhance human cognition.

\section{Related Work}
\subsection{Cognitive function and auditory perception}
Recent studies have showed the potential interplay between cognitive function and auditory perception. %For instance, hearing loss is independently associated with accelerated cognitive decline and cognitive impairment~\cite{lin2013hearing}. Additionally, 
The discussion in~\cite{guinan2006olivocochlear} reveals the mechanical and biological connections between the medial olivocochlear (MOC) efferents in the brain and the auditory system. Additionally, human attention also aids in cocktail party speech perception~\cite{price2021attention, walsh2014selective}. %providing potential to assess cognitive functions through auditory characteristics. 
Another recent study also reported that the selective attention tasks can alter auditory perception in humans~\cite{walsh2014selective}, suggesting that the central nervous system can adjust cochlear micro-mechanics. % before sound signals are converted into neural signals. 
These studies revealed insights into the relationship %provide insights into the role of hearing in attention, and 
between auditory perception and cognitive processes.

%could be a viable approach with significant potential for identifying hearing-related cognitive function assessments. %Despite this evidence, there is no work directly measuring and quantifying the correlation between the human auditory system and cognitive function, especially in real-world settings. 

More in-depth analysis on the active feedback mechanisms of the cochlea also suggest the potential interplay between auditory systems and the brain functions~\cite{kim1986active, davis1983active,hudspeth2014integrating, dang2021joint}. 
While these studies have delineated the biomechanical pathways from the cochlea to the brainstem, they have not directly investigated the relationship between hearing abilities in the human auditory system and cognitive load, nor have they explored the integration with consumer earable sensing in daily listening scenarios.

\subsection{Otoacoustic emissions (OAEs)}
Otoacoustic emissions (OAEs) are commonly used to measure this biomechanism of hearing capabilities, manifesting as sounds that the cochlea emits in response to auditory stimuli,  especially widely used in newborn hearing screenings~\cite{wroblewska2017universal}. They can be detected externally with sensitive microphones placed in the ear canal. Traditional OAEs measurements typically require expensive medical device and a quiet environment, which limits their use to controlled laboratory settings. However, the emergence of wearable sensing technology, particularly ear-worn devices, offers a cost-effective, easy-to-access alternative. This provides the potential to leverage commercially available earbuds equipped with built-in in-ear microphones for OAEs sensing~\cite{shahid2024towards, demirel2023cancelling}. 

Several types of OAEs exist, each with distinct characteristics and stimulus requirements. Spontaneous OAEs (SOAEs) occur without external stimulation; however, they are generally weak and are not present in all individuals, limiting their utility in clinical and research applications~\cite{zurek1981spontaneous}. Distortion product OAEs (DPOAEs) are evoked by simultaneously presenting two primary tones at different frequencies, generating a distortion product at a third frequency within the cochlea~\cite{abdala2001distortion}. Although DPOAEs are widely used for cochlear assessment, most consumer earable devices are equipped with only a single speaker, rendering DPOAE measurements challenging or infeasible in such settings. Furthermore, measuring DPOAE in daily use settings is challenging and intrusive as it requires users to focus on listening to a series of stimuli throughout the assessment~\cite{shahid2024towards}. %\simon{i added a bit, also mention DPOAE is hard to be integrated into daily measurement} 
Single-frequency OAEs (SFOAEs) are elicited by a single tonal stimulus and measure the response at the same frequency, providing an alternative means of assessing cochlear function~\cite{walsh2010properties, quan2025cognitive, guinan2003medial}, with advantages including simpler stimulus delivery and compatibility with single-speaker devices. Work in~\cite{quan2025cognitive} has approved the feasibility of leveraging SFOAEs for assessing cognitive load via the proposed machine-learning-based pipeline. In this study, we adopt SFOAEs as the stimulus for OAE measurement to accommodate the hardware constraints of earable devices while ensuring reliable cochlear assessment.
%\td{@simon, could you please add references for the 2nd paragraph.} \simon{added}

% \td{here add a few sentences on OAEs, including simutabours OAEs which are weak and not present in all particpants, DPOAEs which require two speakers playing different frenquencies while most earables have one, and single frequency OAEs which xx, therefore we adopte SFOAEs as the stimulus for the OAE measurement in this work. Also include the references for OAEs.} \simon{We chose to use a single frequency with Single-frequency OAEs for measurements because of the challenges in playing audio content to elicit cognitive load and playing a stimulus to measure auditory responses simultaneously. This approach aims to avoid altering the audio content for natural perception.}

\section{Study Design and Data Analysis}
\subsection{Study design}
\subsubsection{Overview}
As shown in Figure~\ref{fig:concept}, the earbud speaker plays both the auditory task and the SFOAE stimulus. The auditory tasks, such as bird chirping, human talking, are used to elicit different levels of auditory cognitive load, while the stimulus helps trigger SFOAEs from the ear. These emissions are recorded by the in-ear microphone for analysis. In our study, we represent the auditory event (a wideband signal) as $x_a(t)$, the single-frequency stimulus at frequency $f_s$ as $x_s(f_s, t)$, the OAE response as $s_{\mathrm{OAE}}(f_s, t)$.

\subsubsection{Task Design}
% \begin{figure}
%     \centering
%     \includegraphics[width=1\linewidth]{figs/ACL_Task_Design.pdf}
%     % \vspace{-12pt}
%     \caption{The diagram illustrates the four-level auditory cognitive load task design. The diagram of each task presents a 10-second measurement, where each audio clip contains a 1-second single-frequency stimulus and audio contents are integrated together with the stimulus from Task 2 to 4. To demonstrate the stimulus design, We present a spectrogram view of the last audio clip in Task 4. The audio content spans over a broad frequency band, where the single-frequency stimulus lies in a frequency-cut range spanning from 2.9k to 3.1k Hz. This ensures an interference-free zone to isolate stimuli and OAE responses without being overwhelmed by other audio content.}
%     % \vspace{-12pt}
%     \label{fig:task_design}
% \end{figure}

As shown in Figure~\ref{fig:figure_illustation}, we designed a study comprising four auditory tasks intended to progressively induce increasing levels of auditory cognitive load, drawing inspiration from prior physiological and perceptual research~\cite{walsh2014selective}. Simultaneously, we include the single-frequency stimulus $f_s$ to evoke SFOAE emissions for measurement. Therefore, the input signal delivered through the speaker is:
\begin{equation}\label{eq:input}
 x_i(t) = x_a(t) + x_s(f_s, t)
\end{equation}
In each task, participants listened to a sequence of audio events and have SFOAEs measured, followed by a post-task question to ensure their attention during the task and to support data validity.  %each paired twice with single-frequency tones at 3 kHz, 2 kHz, and 1 kHz. The measurement within each task is repeated twice. From Tasks 2-4, different pre-audio task indications and post-audio questions will be provided in each individual measurement to guarantee participants' attention and active auditory engagement.

Task 1 aims to measure the individuals' auditory responses while they are in a relaxed state without processing any audio sounds, serving as an individual's reference point. To measure the OAEs responses in this relaxed state, we presented only the  single-frequency stimulus, i.e., $x_s(f_s, t)$.% To capture variations in OAEs across frequencies, three single-frequency stimuli at 3 kHz, 2 kHz, and 1 kHz were used. %Each frequency is repeated twice for result validity, resulting in 6 sets of measurements.

Task 2 aims to induce lower-level cognitive load in participants by playing natural sounds devoid of linguistic content. Multiple sound events (e.g., birds chirping, dog barking) are played. 
The single-frequency stimulus is played simultaneously to measure the SFOAEs while participants listen to the audio content. In addition, participants will receive a concise pre-audio prompt (e.g., ‘Please focus on an animal sound.’) to ensure active engagement. Subsequently, a post-audio question (e.g., ‘Did you hear a bird chirping?’) will be administered to confirm the participant’s active perception to guarantee the validity of our acoustic measurement. %In Task 2, 
%Similar as task 1, each sequence consists of three single-frequency tones of 3, 2, and 1 kHz, repeating twice, resulting in 6 sets of measurements. 
%\td{we talk about pre-audio question here but didn't draw it. If you agree with the new structure here, maybe in figure 2 we only keep the b part and include the pre-question in the figure as well, and move the earbuds into next data collection section.} \simon{updated}

Task 3 intends to elevate medium-level cognitive load through language comprehension. In each %1-second 
auditory content clip, there is one person (male or female) reading digits mixed with background noise. The single-frequency stimuli are also played simultaneously for OAEs measurements. The pre-audio prompt guides participants to focus on digits spoken by a specific gender and 
the post-audio question asks participants about the number or order of digits being read (e.g., "What is the last digit spoken by the female?"). %Task 3 contains 12 sets of measurements.

Task 4 is designed to impose a significant auditory cognitive load by simultaneously presenting two spoken digits read by both male and female voices, amidst background noise. Similarly, stimuli are played simultaneously. Participants are tasked with processing the content articulated by both genders and accurately identifying the digits spoken by a specific gender. A pre-audio prompt directs participants to concentrate on either one gender or both, while a post-audio question asks them to recall the digits pronounced by the targeted gender as well as retaining the sequence of the numbers (e.g., ‘What are the last two digits spoken by the male?’). %Task 4 includes 12 sets of measurements.

\subsubsection{Stimulus design} 
To reliably extract the SFOAE response, we analyze the recorded signal from the inward-facing microphone, denoted as $x_o(t)$. This signal contains not only the desired SFOAE component $x_{OAE}(f_s, t)$, but also the playback signal $x_i(t)$, as the speaker and microphone are placed in close proximity. The recorded signal can be expressed as:
\begin{equation}\label{eq:oae}
x_o(t) = x_{OAE}(f_s, t) + x_i(t) = x_{OAE}(f_s, t) + x_s(f_s, t) + \underline{x_a(t)}
\end{equation}

Although our focus is on the SFOAE response $x_{OAE}(f_s, t)$ at the stimulus frequency $f_s$, extracting this component from $x_o(t)$ is challenging due to spectral overlap. For instance, auditory task signals $x_a(t)$ (e.g., bird chirping) span a broad frequency range (typically 1 kHz to 8 kHz), overlapping with the stimulus frequencies (e.g., 3 kHz). As a result, the recorded signal $x_o(t)$ at the stimulus frequency contains the SFOAE response, the stimulus, and unwanted components from the auditory task. While we can control the stimulus $x_s(f_s, t)$ and keep its amplitude constant across sessions, we cannot control the spectral content of $x_a(t)$ at $f_s$ due to the natural complexity and variability of the auditory tasks. This makes isolating the SFOAE response difficult.

To address this, we designed a stimulus modification strategy to minimize spectral overlap and enable reliable SFOAE measurement. Given an audio spectrum shown in Figure~\ref{fig:concept}, we apply a band-stop filter $f_{\mathrm{band}}$ to the auditory task signal $x_a(t)$, removing a narrow 200 Hz frequency band centered at the stimulus frequency $f_s$. The SFOAE stimulus $x_s(f_s, t)$ is then inserted into this vacated frequency band, ensuring minimal interference from the auditory task. The final signal delivered to the participant is:
\begin{align}\label{eq:1}
\hat{x}_i(t) &= x_s(f_s, t) + f_{\mathrm{band}}\left(x_a(t)\right)
\end{align}
leaving $\hat{x}_i(t)$ at $f_s$ equal to only the stimulus $ x_s(f_s, t)$. Therefore, the recorded signal at target frequency $f_s$ is:
\begin{align}\label{eq:filteroae}
   \hat{x}_o(f_s, t) &= x_{OAE}(f_s, t) + \hat{x}_i(f_s, t) \nonumber \\
                     &= x_{OAE}(f_s, t) + x_s(f_s, t)
\end{align}

% \begin{equation}\label{eq:filteroae}
%    \hat{x}_o(f_s, t) &= x_{OAE}(f_s, t) + \hat{x}_i(f_s, t) = x_{OAE}(f_s, t) + x_s(f_s, t)
% \end{equation}
This approach ensures that the SFOAE response remains isolated from the auditory task content, allowing the recorded OAEs at the stimulus frequency to reflect changes in auditory processing related to cognitive load. Importantly, human auditory perception can perceptually "fill in" the removed frequency band~\cite{bregman1994auditory}, maintaining the overall integrity of the auditory task.

\subsection{Data analysis}\label{sec:dataanalysis}
While the primary objective is to analyze SFOAEs under varying levels of cognitive load, it is first essential to validate whether the task design effectively elicits the intended cognitive load to ensure the reliability of SFOAE measurements. The analysis is divided into two parts: an initial verification of the task design through EEG signal analysis and task validation, and an examination of the SFOAE acoustic measurements.

\subsubsection{EEG analysis and task validations}\label{sec:eeganalysis}
%Pre-processing of EEG signals, following~\cite{quatieri2016using}, was imperative to ensure their quality and reliability before further analysis. 
To analyse the EEG signals, we first filter the raw EEG data using a band-pass filter ranging from 1 Hz to 30 Hz. %This filtering aimed to eliminate DC components and encompass the four different frequency bands, including delta, theta, alpha, and beta waves, which are indicative of overall cognitive load. 
Additionally, to counteract power line noise, a notch filter at 50 Hz was employed.
After the filtering process, the data were re-referenced, followed by Independent Component Analysis (ICA) to isolate and remove components associated with artefacts like eye blinks, muscle movements, and cardiac signals~\cite{uriguen2015eeg}.  %using the average reference method. This technique involves using the mean signal from all electrodes as a common reference, thus enhancing signal quality. Independent Component Analysis (ICA) was then applied to isolate and remove components associated with artefacts like eye blinks, muscle movements, and cardiac signals~\cite{uriguen2015eeg}. ICA is known for its effectiveness in artefact removal while preserving the integrity of neural activity.
After pre-processing, the EEG data were divided into segments corresponding to the sessions when audio events were played. This segmentation also synchronized with acoustic measurements from our earable device. The power spectrum was calculated using Fast Fourier Transform (FFT), and the analysis was completed by averaging these power values across measurements and subjects for different tasks. By comparing the EEG power values across tasks, we aimed to determine whether the task design successfully modulates cognitive load, as reflected in neural activity. 

In addition to EEG analysis, we examined participants’ response time and accuracy on the post-audio questions. As task difficulty increases, these behavioral measures reflect cognitive effort and performance. We quantified them by calculating the average response time and average accuracy across participants for each task.
% Following pre-processing, the EEG data were segmented into smaller segments corresponding to the sessions when the audio events were playing to induce the different levels of cognitive load. This segmentation also aimed to synchronize with the acoustic measurements obtained from our earable device. %For each frequency stimulus in a single session, six segments were produced. 
% The power spectrum was calculated after Fast Fourier Transforming (FFT) the signal, and the analysis was finalized by averaging these power energy values for each measurement and subjects for various tasks. %I change the term 'session' to 'measurement', as 'session; does not appear above.

\subsubsection{SFOAE Measurement Analysis}\label{sec_audio_analysis_principle}
We analyze the SFOAE responses to examine their relationship with cognitive load, expecting these responses to vary with different levels of auditory cognitive demand. As described in Eq.~\eqref{eq:filteroae}, our objective is to extract the SFOAE component $x_{OAE}(f_s, t)$ from the recorded signal $x_o(t)$. To achieve this, we first apply a bandpass filter centered at the stimulus frequency $f_s$ to isolate $x_o(f_s, t)$. Next, we perform a fast Fourier transform (FFT) on $x_o(f_s, t)$ to obtain the magnitude at $f_s$, denoted as $M(f_s)$:
\begin{equation}
    M(f_s) = FFT(x_o(f_s, t)) = FFT(x_{OAE}(f_s, t)) + FFT(x_s(f_s, t))
\end{equation}

Since the stimulus $x_s(f_s, t)$ is fixed across tasks and sessions, its FFT component $FFT(x_s(f_s, t))$ remains constant. Therefore, variations in $M(f_s)$ across tasks reflect changes in the SFOAE response.

To account for individual variability, such as differences in auditory perception, ear canal shape, and earbud positioning, we use the magnitude measured during Task 1 as a reference for each participant. The SFOAE magnitudes from Tasks 2 to 4, denoted $M_{i}(f_s)$ for task $i \in \{2,3,4\}$ are compared against this baseline $i=1$. We define the Sound Energy Difference (SED) $\delta_{i}(f_s)$ to quantify acoustic changes for each participant related to cognitive load:
\begin{equation}
   \delta_{i}(f_s) = \left| M_{i}(f_s)^2 - M_{1}(f_s)^2 \right|, \quad i \in [2,4]
\end{equation}
Here, $\delta_{i}(f_s)$ captures how the auditory system response varies under different cognitive loads for each individual. %By analyzing $\delta_{i}$, we assess how OAEs change with increasing auditory task difficulty, as reflected in the stimulus frequency recorded from our earable device.

\section{Data Collection}
\subsection{Earable prototype}
\begin{figure}[t!]
    \centering
    \begin{subfigure}[t]{0.46\columnwidth}
        \centering
        \includegraphics[width=\textwidth]{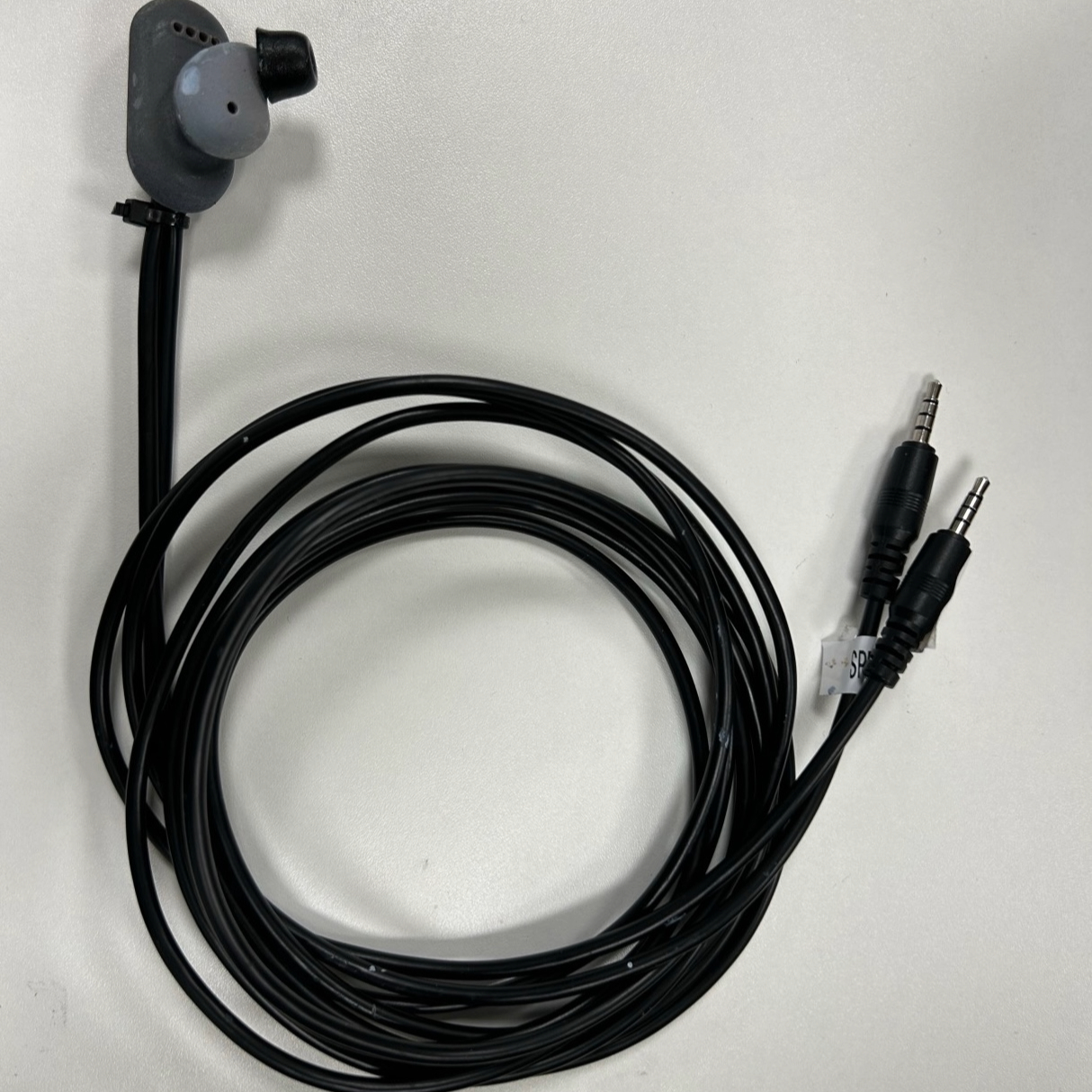}
        \caption{The customized earable prototype with in-ear microphone.}
        \label{fig:earable_device}
    \end{subfigure}
    \hfill 
    \begin{subfigure}[t]{0.46\columnwidth}
        \centering
        \includegraphics[width=\textwidth]{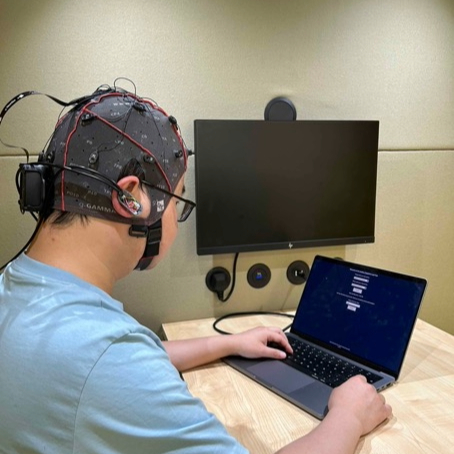}
        \caption{Participants wearing the earables and EEG headset.}
        \label{fig:wear_eeg_sub}
    \end{subfigure}
    \vspace{-8pt}
    \caption{Experimental setup and data collection.}
    \label{fig:experimental_setup}
\end{figure}

To precisely capture SFOAE responses from participants, we designed an earable prototype, an in-ear wearable device, that integrates a 10mm circular compact speaker and an inward-facing analogue microphone with a signal-to-noise ratio of 67 dB(A) (Knowles SPH8878LR5H-1~\cite{KnowlesMicSelection}), shown in Figure~\ref{fig:earable_device}. The microphone and speaker are tightly enclosed by a 3D-printed casing, creating a sealed acoustic chamber that ensures that OAEs reflected from the ear canal can be captured by the microphone. The speaker delivers the auditory tasks and stimuli, while the in-ear microphones simultaneously record SFOAEs for cognitive load analysis, as shown in Figure~\ref{fig:concept}.

\subsection{Data collection}
We collected data from 19 participants, including 11 males and 8 females, aged 20 to 55, with an average of 32, and 42.11\% are female. In total, 192 audio events with stimuli were played, and 30 questions were administered during Tasks 2 to 4. The diverse age range and relatively balanced gender distribution ensure that the study’s findings are contextualized across different demographic groups. Regarding the EEG measurement, 10 of them have given consent to collect EEG data while conducting the study, used for task validation. We used our customised earable prototype and developed an App\footnote{Source code will be open-sourced.} to interact with participants for OAE acoustic measurement, while a 16-channel wearable EEG headset of g.Nautilus RESEARCH\footnote{\url{https://www.gtec.at/product/g-nautilus-research/?srsltid=AfmBOorQABiaUM35quqKvm2uwI6xIeHwi6exGxSoyDhamF5ugOwFpmVe}} is used for EEG data collection. Participants are advised to maintain a stationary sitting posture to minimize extraneous bodily movements which may introduce sound artefacts that would contaminate the data. %\td{we can include a data collection figure with one use wearing the eeg headset. This can make it clear that we use eeg for task validation.} \simon{added}

% \subsection{Earable sensing}\td{@Simon, to complete, can refer to the other paper we have.}
% briefly start with other modalities and applications, and less on the interplay of the brain. Hearing using OAE have been studies, but still limited on hearing loss but not on cogntive load. 

\section{Results}\label{results}
% \begin{figure}[ht!]
%     \centering
%     \includegraphics[width=0.9\textwidth]{figs/overallfigure.pdf}
%     \caption{Overall study design. (a) It consists of four tasks to increase the cognitive load. (b) The earbud prototype. (c) The data collection setup.} %\sw{may consider update a picture with the OmniBuds}\am{I don't think we need the two plots about age range and age groups, we should use the space for something more useful.}\sw{agree, let's put one omnibuds, then finalise something else, then I will update this figure.}\td{for NPJ, I'd suggest we showing data at the beginning, especially this is what we collected instead of public available data. To save space, we may integrate a table showing the data demographic and distributions, kind of a simplified version of this: https://www.nature.com/articles/s41746-024-01136-2}}\sw{\faExclamationCircle~ To-do: Let's finalise the first figure of what should be included. And show a earable prototype picture.}
%     \label{fig:Overall_Figure}
% \end{figure}

%\subsection{Task design and data}\td{@Simon, to reduce. maybe combine with data collection to save space?}\sw{indeed, the design and data collection sections should be allocated before section data analysis}

\subsection{Task design validation}
We investigated the effectiveness of our task design by analysing the EEG signals and post-audio answers across tasks, as shown in Table~\ref{tab:eeg_task_validation}.
\begin{table*}[t]
\centering
\vspace{-8pt}
\caption{Task validation. All EEG power measurements (Total, Delta, Theta, Alpha, Beta) are in $k\textmu V^2$. Each data cell shows the mean$\pm$std with confidence interval.}
\setlength{\tabcolsep}{5pt}
\begin{tabular}{cccccc|cc}
\toprule
Task & Total EEG & Delta (0.5-4 Hz) & Theta (4-8 Hz) & Alpha (8-13 Hz) & Beta (13-30 Hz) & Time (min) & Acc (\%) \\
\midrule
% The \shortstack command was removed from the multicolumn cell below for vertical centering.
Task 1 & \shortstack{11.8±7.0 \\ $[$6.5, 17.6$]$} & \shortstack{5.4±2.2 \\ $[$2.9, 8.4$]$} & \shortstack{2.9±1.1 \\ $[$1.6, 4.2$]$} & \shortstack{2.3±0.9 \\ $[$1.2, 3.3$]$} & \shortstack{1.2±0.5 \\ $[$0.6, 1.8$]$} & \multicolumn{2}{c}{Baseline / No Q\&A} \\ 
\addlinespace
Task 2 & \shortstack{22.9±4.2 \\ $[$20.3, 26.3$]$} & \shortstack{9.1±0.9 \\ $[$7.9, 10.5$]$} & \shortstack{6.9±0.8 \\ $[$6.0, 8.1$]$} & \shortstack{4.1±0.4 \\ $[$3.6, 4.9$]$} & \shortstack{2.8±0.4 \\ $[$2.2, 3.4$]$} & \shortstack{2.7±0.7 \\ $[$2.2, 3.1$]$} & \shortstack{75.0±18.3 \\ $[$63.2, 86.8$]$} \\
\addlinespace
Task 3 & \shortstack{103.6±22.0 \\ $[$86.2, 117.7$]$} & \shortstack{35.9±4.5 \\ $[$28.5, 41.3$]$} & \shortstack{26.3±3.8 \\ $[$20.6, 31.7$]$} & \shortstack{20.9±3.1 \\ $[$16.1, 25.0$]$} & \shortstack{20.5±2.4 \\ $[$17.2, 24.7$]$} & \shortstack{5.2±0.9 \\ $[$4.6, 5.8$]$} & \shortstack{73.3±11.8 \\ $[$65.7, 80.9$]$} \\
\addlinespace
Task 4 & \shortstack{299.9±30.3 \\ $[$273.9, 317.3$]$} & \shortstack{90.8±7.5 \\ $[$78.5, 101.9$]$} & \shortstack{59.6±4.9 \\ $[$50.7, 68.2$]$} & \shortstack{73.2±5.2 \\ $[$62.3, 78.8$]$} & \shortstack{76.3±4.7 \\ $[$66.9, 83.0$]$} & \shortstack{5.4±0.9 \\ $[$4.8, 5.9$]$} & \shortstack{69.2±14.4 \\ $[$59.9, 78.5$]$} \\
\bottomrule
\end{tabular}
\label{tab:eeg_task_validation}
\vspace{-9pt}
\end{table*}
The EEG power spectral analysis demonstrated increasing EEG energy associated with incremental cognitive load across Tasks 1 to 4, consistent with findings reported in \cite{quan2025cognitive}. Detailed analysis of the four EEG frequency bands, delta, theta, alpha, and beta, also revealed a similar increasing trend, further validating the effectiveness of the task design. Additionally, analysis of response times and accuracy on the post-audio questions showed that as task difficulty increases, participants take longer to respond, however, receiving worse accuracy performances. This further confirms that our task design effectively induces increasing cognitive load across tasks elicited by our designed auditory events.

\subsection{SFOAEs with increasing cognitive load}%\td{Simon, to reduce.}
% \subsubsection{Acoustic measurements across tasks.}
\begin{figure}[!t]
    \centering
    \includegraphics[width=\linewidth]{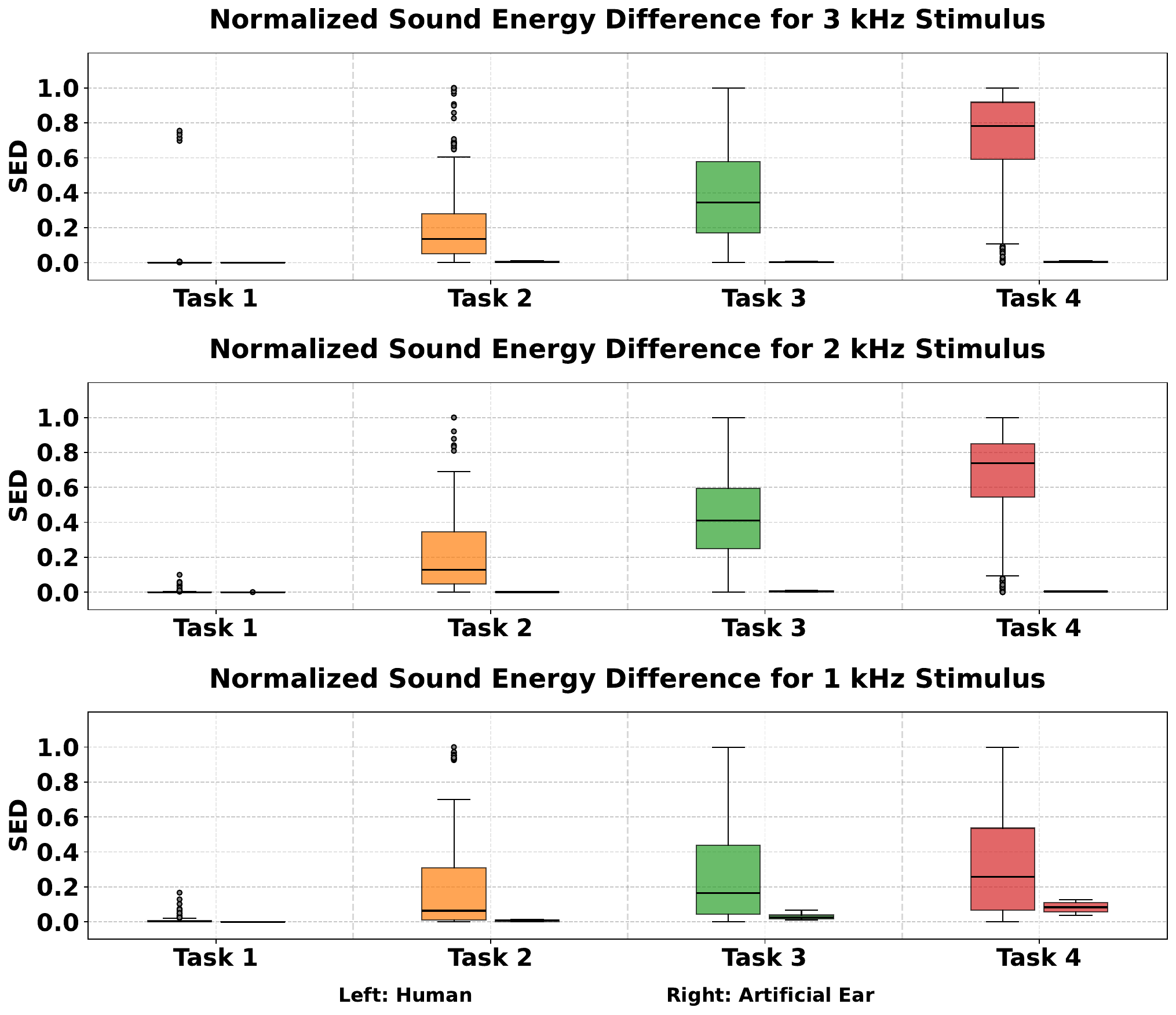}
    \caption{Comparison between sound energy density (SED) from the (left) human ear and (right) artificial ear at 3kHz, 2kHz and 1kHz respectively.}
    \vspace{-8pt}
    \label{fig:human_vs_dsp}
\end{figure}
To determine if increasing cognitive load induces specific changes in the SFOAEs captured by an earable device, %. Using Task 1 as a baseline for each individual, 
we compared the Sound Energy Differences (SED) (refer to Section~\ref{sec_audio_analysis_principle}) from Tasks 1 to 4. %for Tasks 2 to 4, which involved progressively higher cognitive loads, relative to the acoustic responses from Task 1. The acoustic responses are represented and compared in terms of sound energy difference (SED) (refer to Section~\ref{sec_audio_analysis_principle}) %, which quantifies the changes in OAEs for each task relative to the baseline task 1. 
As shown in Figure~\ref{fig:human_vs_dsp}, an increasing trend is observed from Task 1 to Task 4, and this trend remains consistent across various frequency measurements at 3k, 2k, and 1k Hz, with high significance (p $<$ 0.01) for all participants. These findings indicate that increasing cognitive load significantly alters hearing characteristics consistently and can be captured via the earable device. 

In addition, we also compared acoustic measurements from human ears to those obtained from an artificial ear (MiniDSP Ears\footnote{\url{https://www.minidsp.com/products/acoustic-measurement/ears-headphone-jig}}), which has no neural feedback mechanisms as human does. The artificial ear exhibited no significant acoustic response variations across all tasks under different frequency stimuli. This comparison results suggest that changes in acoustic responses observed in human ears are inherently derived from neural feedback. This further validates the interplay between neural responses and auditory perception,  demonstrating the potential of earable sensing technologies for inferring cognitive load.

\subsection{SFOAE variations across stimuli frequencies}%\td{Simon, to reduce.}
To explore how cognitive load influence hearing capabilities across different frequencies, we further analysed the SEDs at different frequencies for each individual separately. %We propose to assess the sensitivity to each frequency stimulus for each participant, by analyzing the trend in SED responses from Task 1 to 4. 
We quantify the trend of SEDs from Task 1 to 4, using an Ordinary Least Squares (OLS) linear regression model, where a higher regression coefficient indicates a rapid change from Task 1 to 4, thus reflecting a high sensitivity of auditory perception to alterations caused by increasing cognitive load, and vice versa.
\begin{figure}[h]
    \centering
    \includegraphics[width=1\linewidth]{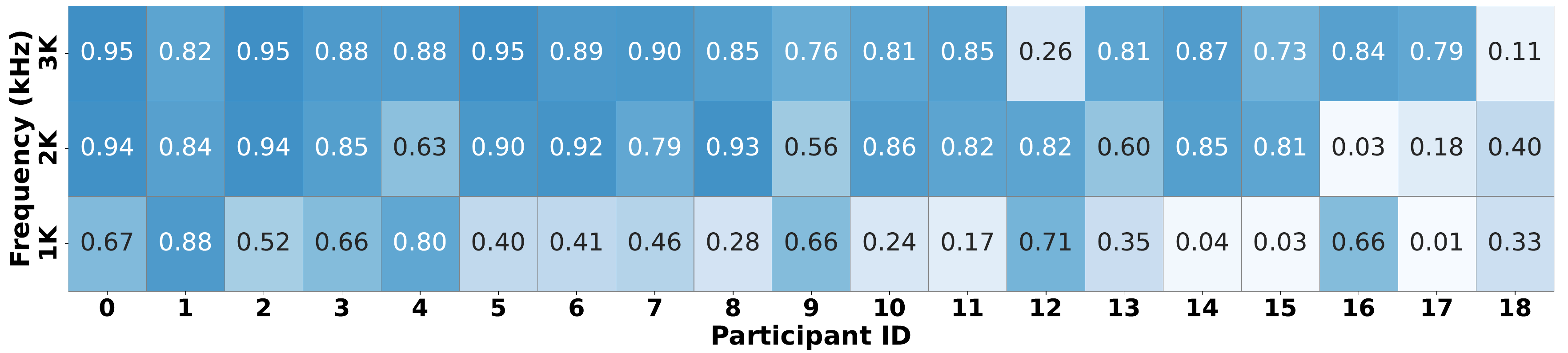}
    \vspace{-8pt}
    \caption{The quantification of the trends in terms of sensitivity for frequencies at 3 kHz, 2 kHz, and 1 kHz across all participants. A higher value indicates a significant increase in SEDs from Task 1 to 4, suggesting greater sensitivity.}
    \label{fig:het}
\end{figure}
\begin{figure}[h]
    \centering
    \includegraphics[width=\columnwidth]{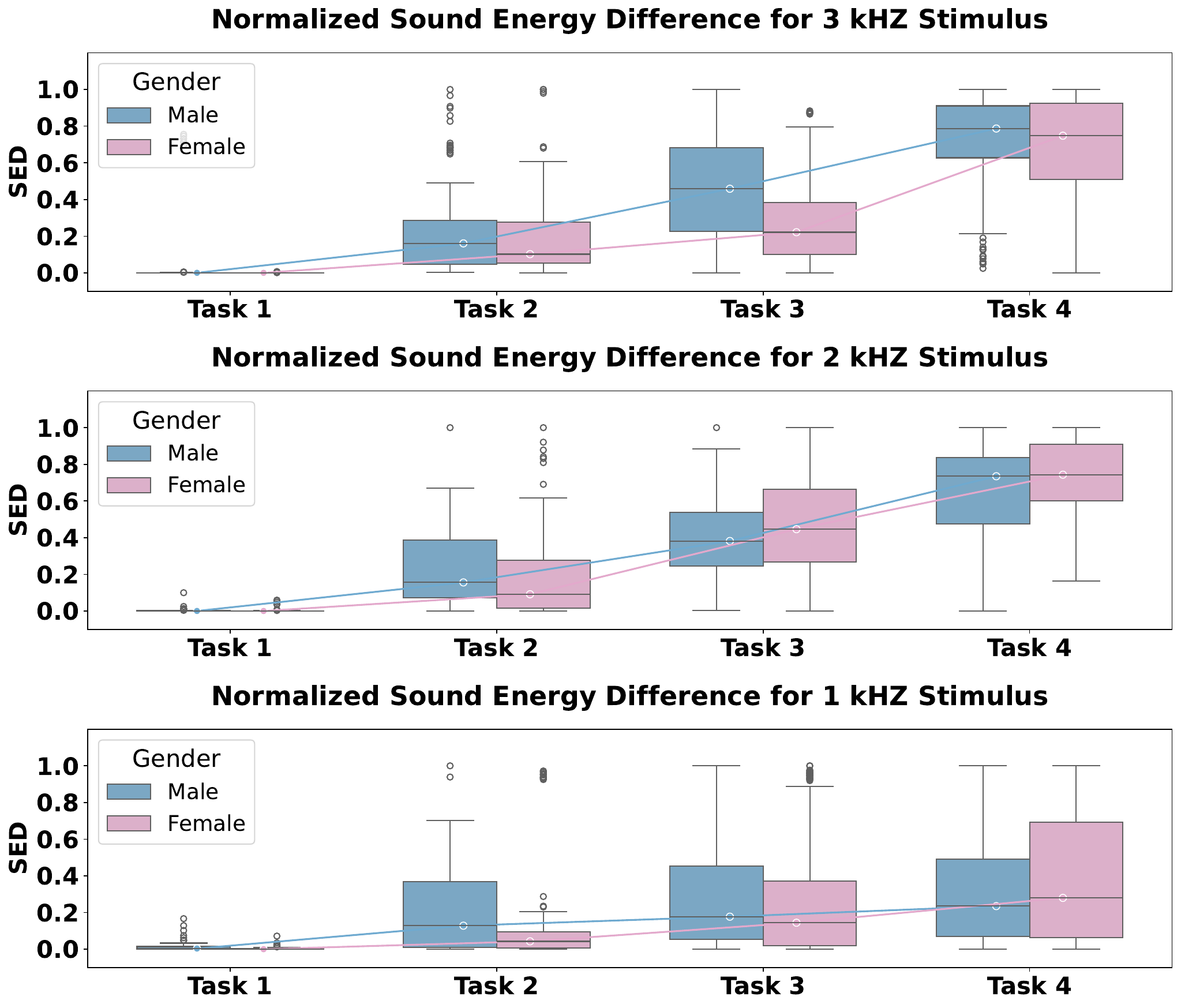}
    \vspace{-8pt}
    \caption{Comparison of SFOAEs across gender groups in terms of SED.}
    \label{fig:trend_by_gender}
\end{figure}
% \begin{figure}[!t]
%     \centering
%     \includegraphics[trim=25 0 100 0, clip, width=\linewidth]{figs/linear_model_heatmap.pdf}
%     \caption{The quantification of the trends in terms of sensitivity for frequencies at 3 kHz, 2 kHz, and 1 kHz across all participants. A higher value indicates a significant increase in SEDs from Task 1 to 4, suggesting greater sensitivity.\td{make font size larger?}\sw{updated.}}
%     \label{fig:het}
% \end{figure}
Figure~\ref{fig:het} presents the regression coefficients for all individuals under three frequency stimuli: 3k, 2k, and 1kHz. We observed that individuals exhibited the greatest sensitivity at 3 kHz, followed by 2 kHz, and showed the least sensitivity at 1 kHz, with average coefficients across participants of 0.87$\pm$0.06 (12 participants), 0.79$\pm$0.18 (6 participants), and 0.88 (only 1 participant), respectively. While most individuals commonly exhibit this sensitivity trend, they display different levels of sensitivity, and some show varied sensitivity across different frequencies. %For instance, individual 12 exhibits the highest sensitivity at 2 kHz and the lowest at 3 kHz. 
This heterogeneity among individuals suggests potential variations in subjective hearing capabilities and hearing sensitivity, the fit of the earable devices, and the diverse mechanisms of neural processing for auditory inputs.

\subsection{Impact of gender and age}
%Additionally, we aimed to understand whether the acoustic earable sensing of cognitive load exhibits different patterns across gender and age, due to their varying auditory perception capabilities. 
To understand whether the sfOAEs exhibits different patterns across gender and age, we compared SEDs across female and male groups and among three age groups, from Tasks 1 to 4.
% \begin{figure*}[!htbp]
%     \centering
%     \begin{subfigure}{0.45\textwidth}
%         \centering
%         \includegraphics[width=\textwidth]{figs/all_participants_trend_by_gender_with_trend_lines.pdf}
%         \caption{Trend by Gender}
%     \end{subfigure}
%     \hfill
%     \begin{subfigure}{0.45\textwidth}
%         \centering
%         \includegraphics[width=\textwidth]{figs/all_participants_trend_by_3age_group_with_trend_lines.pdf}
%         \caption{Trend by Age Group}
%     \end{subfigure}
%     \caption{Comparison of trends among different participant groups.}
%     \label{fig:trend_by_groups}
% \end{figure*}

% \begin{figure}[!t]
%     \centering
%     \includegraphics[width=0.88\linewidth]{figs/subgroups_comparison.pdf}
%     \caption{Comparison of SED results within in the ages subgroups (left) and the gender subgroups (right).}
%     \label{fig:age_and_gender_sed}
% \end{figure}

\subsubsection{Gender differences}
Figure~\ref{fig:trend_by_gender} indicates no apparent gender bias across all three frequencies, as both female and male participants show similar acoustic measurement responses. The trend is more pronounced at 3 kHz than at 2 kHz and 1 kHz, consistent with previous findings. Interestingly, at 3 kHz, males show a greater increase in response than females, possibly suggesting a slight difference in perceiving high-frequency sounds under varying cognitive loads. This aligns with previous research~\cite{fasanya2012gender}, which found that males outperform females in detecting changes in auditory signals.
% Figure~\ref{fig:trend_by_groups}(a) shows that gender bias is not apparent across all three frequencies, as both female and male participants exhibit similar responses in terms of both the mean and range of acoustic measurements. The overall trend at 3 kHz is more pronounced than at 2 kHz and 1 kHz, aligning with previous findings for all participants. Interestingly, at 3 kHz, male participants tend to show a greater increase in response compared to female participants,  potentially suggesting a slight difference in perceiving sounds at a relatively high frequency under different cognitive loads compared to females.
% This finding shares commonality with previous research~\cite{fasanya2012gender}, which demonstrated that males outperformed females in detecting changes in auditory signals. %While our task is designed to elicit different levels of cognitive load, it also presents changes in the auditory stimulus from Task 1 to 4.
%\vspace{-10pt}
\subsubsection{Age differences}
\begin{figure}[]
    \centering
    \includegraphics[width=\columnwidth]{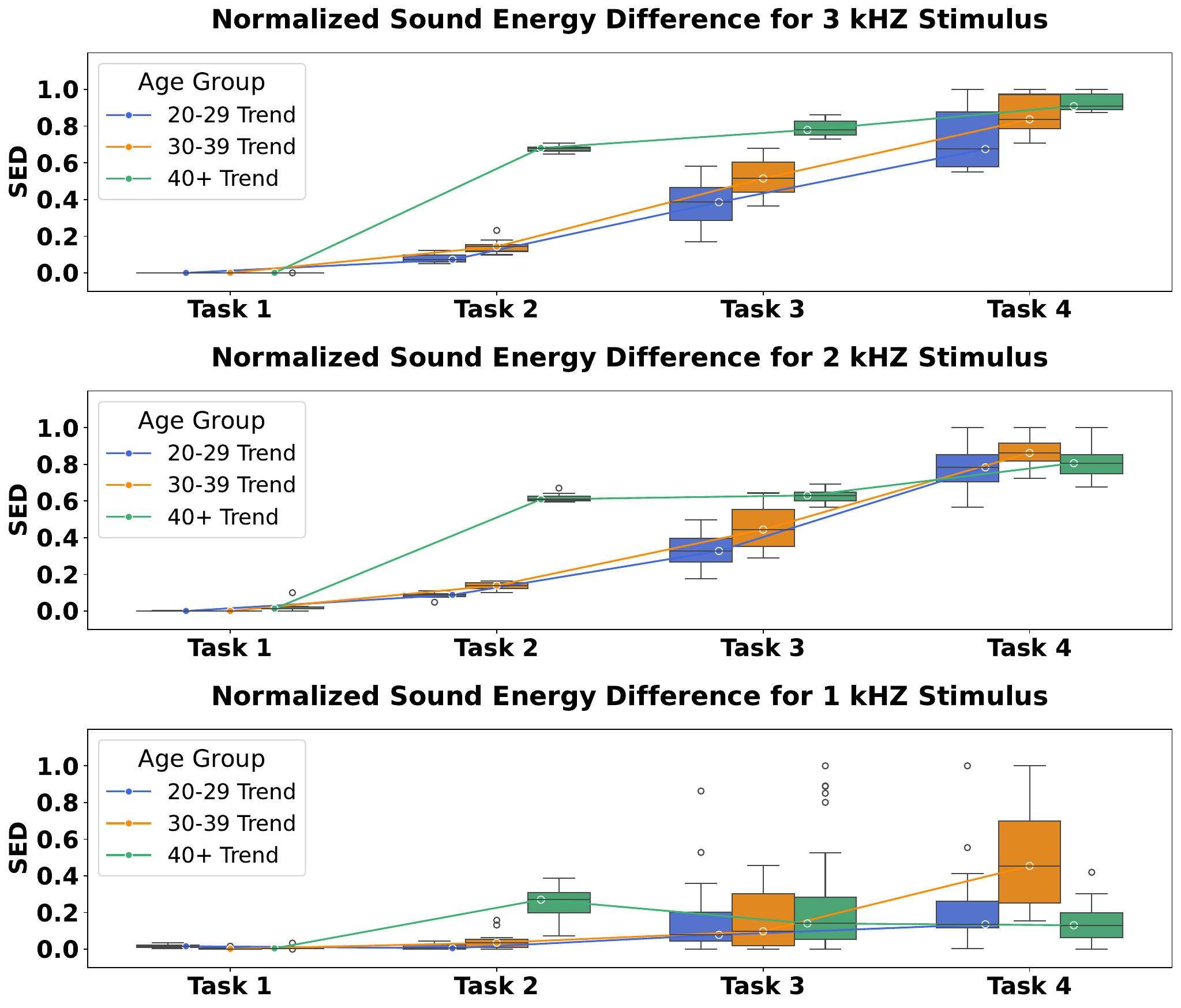}
    \caption{Comparison of SFOAEs across age groups in terms of SED.}
    \label{fig:trend_by_age}
    \vspace{-10pt}
\end{figure}
Figure~\ref{fig:trend_by_age} shows a similar response trend for the 20-29 and 30-39 age groups at 2 kHz and 3 kHz, but a significant increase from Task 1 to 2 for the 40+ age group, indicating differences in perceiving acoustic events likely due to age-related hearing decline. Age-related perceptual declines in hearing and vision generally begin in the fourth decade, followed by cognitive decline~\cite{pichora2006effects, baltes2004wisdom}, which might explain the increased effort required by the 40+ age group to process auditory stimuli. There is no conclusive evidence at 1 kHz consistent with Figure~\ref{fig:human_vs_dsp}, as the trend is generally less pronounced and may require further investigations.

\section{Discussion and Conclusion}\label{discussion}
Our study is pioneering in leveraging earable acoustic sensing to infer cognitive load by examining changes in auditory characteristics using SFOAEs. Specifically, we found that 63.2\% of participants (12 out of 19) exhibited the greatest auditory sensitivity at 3 kHz under varying cognitive loads, with a minimum sensitivity value of 0.76. This suggests that 3 kHz could potentially serve as a key frequency for cognitive load screening. Additionally, we observed variations across different demographic groups, highlighting individual differences. Furthermore, we present a proof-of-concept demonstrating that commercially available in-ear wearable devices can, to some extent, support cognitive load screening. This lays the foundation for scalable, population-wide assessment of cognitive states using acoustic sensing.

We also acknowledge that the current measurements are limited by the use of discrete frequency tones within a narrow frequency range. To enable a broader and more robust characterization of cochlear and cognitive dynamics, future work will explore the use of broadband or frequency-swept stimuli embedded within natural audio content. This approach would not only allow for continuous and unobtrusive monitoring but also expand the measurable frequency range beyond isolated tones. In addition, validation on larger and more diverse cohorts will be essential to generalize the findings and assess the feasibility of cognitive state monitoring at scale using earable acoustic sensing.

Overall, this work demonstrates the feasibility and promise of using earable acoustic sensing as a non-invasive and scalable method for inferring cognitive load through auditory system dynamics. By leveraging SFOAEs and off-the-shelf earables, we provide early evidence that in-ear devices can move beyond traditional hearing assessments to support cognitive state monitoring in real-world environments. This study lays important groundwork for future development of personalized, passive, and continuous cognitive sensing technologies that are both accessible and embedded in everyday auditory interactions.

% Future work will focus on using sweeping sounds that span the entire spectrum to elicit responses while embedding these sounds within the audio content, aiming to replace the single-tone stimulus for more comprehensive measurement while making the sounds imperceptible during natural interactions.
% %Additionally, while we have used various ear tip sizes to accommodate different ear canal dimensions, a more advanced method for quantifying and calibrating earable fitting—beyond subjective perception—could be implemented to standardize signal quality across individuals.

% \section{Acknowledgments}

% Identification of funding sources and other support, and thanks to
% individuals and groups that assisted in the research and the
% preparation of the work should be included in an acknowledgment
% section, which is placed just before the reference section in your
% document.

% \begin{acks}
% \end{acks}

\bibliographystyle{ACM-Reference-Format}
\balance
\bibliography{ref}

% %%
% %% If your work has an appendix, this is the place to put it.
% \appendix









\end{document}